\documentclass[conference]{IEEEtran}
\pdfoutput=1

\usepackage[utf8]{inputenc} 
\usepackage[english]{babel}
\usepackage[T1]{fontenc}  

\usepackage{epsfig}           
\usepackage{subfigure}
\usepackage{graphicx}
\usepackage{setspace}
\usepackage{amsmath}
\usepackage{amsfonts}
\usepackage{amssymb}
\usepackage{calligra}
\usepackage{eucal}
\usepackage[linesnumbered]{algorithm2e}
\usepackage{todonotes}
\usepackage{multirow}

\usepackage[hyphens,spaces,obeyspaces]{url}
\urldef{\footurl}\url{http://www.inf.ufpr.br/lesoliveira/download/modelo_w2v_vec600_wd10_ct5_tec1.txt.zip}

\usepackage{lscape}

\graphicspath{{Figures/}}

\IEEEoverridecommandlockouts

\begin {document}

\title{Legal Document Classification: An Application to Law Area Prediction of Petitions to Public Prosecution Service}

\author{
     \IEEEauthorblockN{Mariana Y. Noguti}
 	\IEEEauthorblockA{MPPR, DInf/UFPR, Brazil\\
 		mynoguti@mppr.mp.br}
 	\and
 	\IEEEauthorblockN{Eduardo Vellasques}
 	\IEEEauthorblockA{SAP SE, Germany\\
 		eduardo.vellasques@sap.com}
 	\and
     \IEEEauthorblockN{Luiz  S. Oliveira}
 	\IEEEauthorblockA{DInf/UFPR, Brazil\\
 		luiz.oliveira@ufpr.br}
 	}
\maketitle


\begin{abstract}
    In recent years, there has been an increased interest in the application of Natural Language Processing (NLP) to legal documents. The use of convolutional and recurrent neural networks along with word embedding techniques have presented promising results when applied to textual classification problems, such as sentiment analysis and topic segmentation of documents. This paper proposes the use of NLP techniques for textual classification, with the purpose of categorizing the descriptions of the services provided by the Public Prosecutor's Office of the State of Paraná to the population in one of the areas of law covered by the institution. Our main goal is to automate the process of assigning petitions to their respective areas of law, with a consequent reduction in costs and time associated with such process while allowing the allocation of human resources to more complex tasks. In this paper, we compare different approaches to word representations in the aforementioned task: including document-term matrices and a few different word embeddings. With regards to the classification
    models, we evaluated three different families: linear models, boosted trees and neural networks. The best results were obtained with a combination of Word2Vec trained on a domain-specific corpus and a Recurrent Neural Network (RNN) architecture (more specifically, LSTM), leading to an accuracy of 90\% and F1-Score of 85\% in the classification of eighteen categories (law areas).
\end{abstract}
    

\section{Introduction}

The Public Prosecutor's Office of the State of Paraná (\textit{Ministério Público do Estado do Paraná - MPPR}), Brazil, is an institution responsible for representing the interests of society, acting directly in several areas related to the fundamental rights of citizenship, such as the defense of public health, environment, public patrimony, human rights, among others. \footnote{In Brazil, the Public Prosecution Office acts as both a Public Defender's Office and as the Public Prosecutor's Office itself} One of its main duties is to receive the petitions of the general population through onsite appointment, phone calls and anonymous reports, and subsequent forwarding to the most appropriate sector, being essential for this task the identification of the field of law related to the case. Thus, a request for access to a new medicine, for example, should be sent to the Health Prosecutor's Office, while a request related to daycare vacancies should be directed to the education board. Currently, the MPPR has about 470 units throughout the state of Paraná that receives, on average, more than ten thousand requests per month. All information is registered in an electronic system called PRO-MP and validated by a Prosecutor.

A sample of 24,532 calls revealed that approximately 28\% of the records analyzed presented inconsistent information, usually due to the mismatch of the petition (a short text) and the area of law (categorical variable chosen by the user). The wrong association can be explained due to the fact that most of the employees involved lack a proper legal training, making the task even more challenging. In addition, the absence of an unified protocol for filling in information in the system reduces the uniformity and precision of the registered data. This has a negative impact in the decision making process (made by senior management), as law area is often used to determine the allocation of human and material resources in areas with higher demand. Moreover, the inaccuracy in filling the area of law may lead to requests being forwarded to wrong units, increasing its overall processing time.

Considering the above, in this paper we investigate different classification methods and representation techniques to automatically predict the area of law of petitions to the Public Prosecutor's Office. The rationale behind this is to reduce the interpretative variability of the cases, as well as the time spent by the user in the selection of the area of law in the system, allowing the allocation of this time in more complex tasks. In addition to the immediate gain at the time of registration, an appropriate classifier would allow to remove the validation step from the State Prosecutor, automatically forwarding the petition to the appropriate sector, thus reducing the time to process cases. Finally, the continued use of the proposed method will make it possible to generate more reliable statistics on population needs and consequently, make Public Prosecution more efficient.

In the following sections, we rely on a dataset obtained from the PRO-MP system regarding the registration of the public petitions. The main contributions of this paper are: (1) in terms of word representation, we demonstrate that simpler approaches like TF-IDF benefit from more extensive pre-processing, while semantic-oriented approaches like word embeddings can be applied with minimal pre-processing, (2) we compare the effectiveness of embeddings trained on generic documents against domain-specific embeddings and identified improvements in the latter (trained on Brazilian Portuguese legal documents), (3) we provide a new set of word embeddings pre-trained on Brazilian Portuguese legal documents for future applications on the same field of research.\footnote{\footurl} (4) finally, we provide some best practices and recommendations for the use of text classification as a way to make public service more efficient, demonstrating that professionals in this field would greatly benefit from the automation of tasks like the one presented here.

Particularly in the empirical study, we evaluated a variety of classification models, having identified better performance in the use of Recurrent Neural Networks (RNN) to classify the area of law of short texts. The representation obtained with domain-specific Word2Vec \cite{Mikolov-2013} applied alongside with the Long Short-Term Memory (LSTM) \cite{Hochreiter-1997} RNN architecture achieved the best results, with an accuracy of 90\% and F1-Score of 85\% in the task of classifying eighteen different fields of law.

The rest of this paper is organized as follows. In Section \ref{sec:related} we present some related research work, in Section \ref{sec:method} we present the proposed method. Experimental results and discussion are presented in Section \ref{sec:results}. Finally, Section \ref{sec:conclusion} concludes this paper and points to some future directions.

\section{Related Work}
\label{sec:related}

Natural Language Processing (NLP) has gained a lot of popularity in the legal community in recent years, motivated especially by the potential use of unstructured data contained in various documents and records and the recent progress in research in this field. Particularly, in the legal area, the application of machine learning techniques and other technologies is relatively recent, notwithstanding, has led to the emergence of Legal Tech companies, which propose the convergence of legal practice and technology.

With regard to the research community, we can point out the work of Sulea et al. (2017) \cite {Sulea-2017}, which used court decisions by the Supreme Court of France along with various metadata to predict the law area of the analyzed cases, among other tasks. For this problem, they used ensembles of classifiers combined with simple word representations (mostly uni-grams and bi-grams). In a line of research more similar to ours, Undavia et al. (2018) \cite {Undavia-2018} proposed to classify US Supreme Court documents into fifteen different categories, comparing various combinations between feature representation and classification models, with better results through the application of Convolutional Neural Networks (CNN) to Word2Vec \cite{Mikolov-2013} representations. A similar model was applied to documents received by the Supreme Court in Brazil and reported in Silva et al. (2018) \cite {Silva-2018}, which obtained a classifier using an embedding layer and a CNN applied to a problem involving six classes.

The recent literature shows an increasing interest in the use of different RNN architectures, such as Long Short-Term Memory (LSTM) \cite{Hochreiter-1997} and Gated Recurrent Unit (GRU) \cite{Chung-2014} to solve NLP problems. Wang et al. (2018) \cite {Wang-2018} presented a comparative study between LSTM and other classifiers applied to sentiment analysis in short texts. Three different datasets were tested, with better performance using LSTM in cases where there was a large volume of data in the training partition. In another study, Yin and Kann (2017) \cite {Yin-2017}, did a comparison between CNN and RNN structures applied to seven different tasks. The study found that RNNs performed well in a variety of NLP related tasks. Still on the subject, Wang et al. (2016) \cite {Wang-2016} compared LSTM and GRU (associated with CNNs) for three different problems of text classification. They reported that RNN associated with CNN surpassed the other classification strategies they have considered in their study.

\section{Proposed Method}
\label{sec:method}

In this section we describe the methodology used to tackle law area classification. It follows the classical text classification methodology with dedicated modules for pre-processing, feature extraction, and classification. 

\subsection{Pre-processing}

Our pre-processing pipeline consists of (1) tokenization (dividing sentences into words), (2) lowercasing (standardizing letters to lowercase), (3) punctuation removal, (4) Part-of-Speech (PoS) tagging, (5) numeral normalization (transforming numerals to zero) (6) name normalization (mapping proper names to the \textit{proper\_name} token), based on a registry of Brazilian Portuguese names collected by the Brazilian Institute for Geography and Statistics \cite{IBGE-2010}, (7) lemmatization to reduce the variability caused by inflection (given that Portuguese is a highly inflected language), and (8) non-ASCII character removal.

To perform word normalization, we applied some custom-designed regular expressions, while the lemmatization process and PoS tagging were based on what is available in the spaCY \footnote{https://spacy.io} library (for Portuguese language). It is noteworthy that the pre-processing steps were applied differently according to the selection of the feature representation technique (as word embeddings and document-term matrix entail different pre-processing methods). We describe the specific details in the next section.

\subsection{Feature Extraction}

After pre-processing, the next step consists of converting the text into an appropriate representation. TF-IDF and word embeddings are two of the most common approaches in NLP to represent words. However, given that those two representations are fundamentally different (in the former words are represented by a compatibility score between terms and documents while in the later words are represented as a vector of real values), we decided to create two distinct pre-processing modules as described below:

\begin{enumerate}
\item Complete: composed of all the pre-processing steps described in the previous section.
\item Simple: composed of tokenization, lowercasing, replacing URL and email structures into the tokens "URL" and "EMAIL" and standardization of numerals to the number zero.
\end{enumerate}

Considering that the application of TF-IDF generally requires greater word standardization to group corresponding tokens, we have used the complete aforementioned pre-processing to reduce the variability of terms found in the description texts. In the PoS tagging stage, we filtered some classes of words, keeping only the following classes -- noun, verb, adverb and adjective -- as those tend to be the most relevant for detecting a subject in the petition description. This step reduces the size of the vocabulary and also has the effect of stop word removal as it excludes articles from the texts.

After that, we created a TF-IDF for both uni-grams and bi-grams. We only considered tokens that have a frequency greater than ten in the corpus, removing typos and very rare words. We also removed stop-words (frequency greater than 90\%), as they have no impact in distinguishing the different classes. For simplicity, we call this feature extraction method ``TF-IDF Small''.

In the case of word embeddings, we used the ``Simple'' version of the pre-processing since the embedding techniques are capable of extracting semantic relationships between words, projecting them into (semantically) meaningful vector space and requiring less clean up tasks \cite {Chen-2013}. We employed Word2Vec, trained on a set of legal documents. For the sake of comparison, we also employed pre-trained (generic) embeddings. More details about the training process are provided in Section \ref{sec:results}.

\begin{figure}[h]%
  \centering
  \includegraphics[width=8cm]{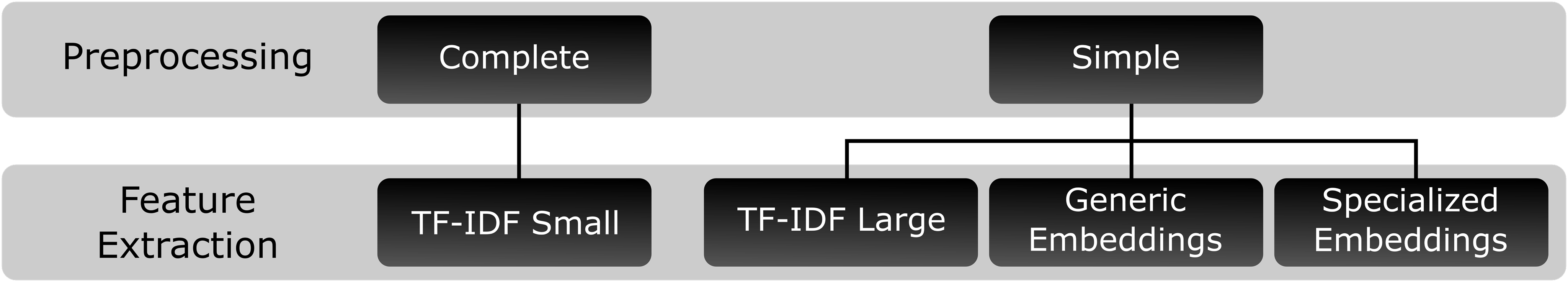}
  \caption{Methodology applied to the dataset}
  \label{fig:methods}
\end{figure}

To better compare the impact of the pre-processing we have used, as mentioned before, a ``Simple'' pre-processing in the construction of a TF-IDF matrix as well. In this case we also created a TF-IDF matrix with uni-grams and bi-grams and configured the tokens to be considered to have more than ten counts and less than 90\% frequency in the corpus. This feature extraction method is called ``TD-IDF Large''. Figure \ref{fig:methods} summarizes the pre-processing and feature extraction techniques considered in this paper.

\subsection{Classifiers}
As mentioned before, in this study we have considered three different families of classification models: linear (logistic regression and support vector machines, with both RBF and linear kernels), boosted trees (random forest and gradient boosting) and neural networks (text CNN and gated RNNs).

With the TF-IDF matrices (small and large), linear classifiers and boosted trees are employed while in the case of word embeddings (generic and specialized), all three categories of classifiers are considered. For the first two groups of classifiers, the sentence representation is obtained by averaging all word embeddings that compose the sentence whereas in the application of neural networks we used an embedding layer, padding the sentence size to 300. It is worth remarking that pre-trained embeddings are used in the embedding layer as a lookup table (not fine-tuned for the new classification task). We relied on Keras\footnote{https://keras.io/} for the experiments with neural networks (using a TensorFlow\footnote{https://www.tensorflow.org/} backend).

It is important to note that we have not made use of TF-IDF matrices in conjunction with neural networks since in term-frequency representations such as TF-IDF, each sentence is represented as a vector composed of $N$ numeric values, where $N$ is the number of columns in the TF-IDF matrix and each numeric value is a word in the corpus vocabulary. In contrast, in the case of word embeddings, each sentence is represented as a matrix with $N$ rows and $K$ columns, where the rows are the words that compose a sentence and the columns are the elements of the vector that represents each word (in this case, each sentence is represented by a matrix with 300 rows and 600 columns). The last representation is the expected format in the embedding layer of both convolutional and recurrent neural networks \cite{Yin-2017}.
Besides that, specifically in the case of RNNs, the sequence of words that compose the sentence is considered as an important feature for the functioning of these architectures as in this type of neural network the text is analyzed sequentially and the semantics of the previous terms are preserved during the training. Considering that TF-IDF techniques are similar to bag-of-words, the representation of a sentence does not preserve any ordering of the words that compose it, since it is represented as an unordered set of terms \cite{Hassan-2017}, and the use of this kind of representation is not suited for RNNs.

\section{Results and Discussion}
\label{sec:results}

\subsection{Dataset}
\label{sub-sec:dataset}
\begin{figure}[h]%
  \centering
  \includegraphics[width=8cm]{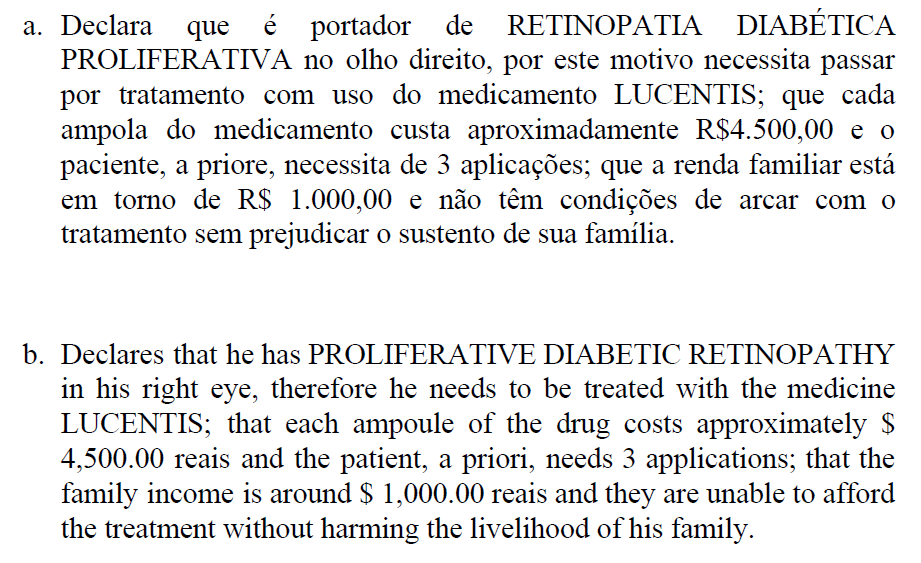}
  \caption{An example of a text extracted from the dataset: (a) Portuguese and (b) English.}
  \label{fig:exampleoftext}
\end{figure}

The dataset considered in this work contains 17,740 documents from 18 different fields of law. It has been obtained from petitions registered  in the PRO-MP system between 2016 and 2019. All the samples of this dataset were carefully reviewed and labelled by prosecutors from the MPPR. Figure \ref{fig:exampleoftext} shows an example of text of the dataset.

Table \ref{tab:classes} details the dataset presenting the number of samples and the average size (in words) for each class. As one may observe, it is a highly unbalanced dataset. Family, Health and Education law are the top-3 largest types of petition.

\begin{table}[!htb]
	\caption {Distribution of the dataset}
	\begin{center}
		\begin{tabular}{l @{\hspace{0.1\tabcolsep}} c @{\hspace{1\tabcolsep}} c} \hline
		    \multirow{2}{*}{Law Area} & Number of & Average \\
		                              & Samples   & Tokens \\ \hline
			
			Child and Youth (CHI)                & 696   & 81 \\
			Civil (CIV)                          & 771   & 48 \\
	        Consumer (CON)                       & 611   & 51 \\
	        Criminal (CRI)                       & 1,060 & 67 \\
	        Disability Rights (DIS)              & 589   & 80 \\
	        Domestic Violence (DOM)              & 419   & 63 \\
	        Education (EDU)                      & 1,237 & 74 \\
	        Elder (ELD)                          & 411   & 133 \\
            Electoral (ELE)                      & 267   & 53 \\
            Environmental (ENV)                  & 351   & 62 \\
            Family (FAM)                         & 5,995 & 42 \\
            Health (HEA)                         & 2,935 & 68 \\
            Human Rights (HUM)                   & 651   & 37 \\
            Labor (LAB)                          & 256   & 45 \\
            Misconduct in Public Office (MIS)    & 415   & 76 \\
            Registration (REG)                   & 430   & 43 \\
            Social Security (SOC)                & 256   & 44 \\
            Urban Planning (URB)                 & 390   & 83 \\ \hline
            
			\label{tab:classes}
		\end{tabular}
	\end{center}
	\vspace{-5mm}
\end{table}

In this work we have split the dataset into 90\% and 10\% for training and testing, respectively. 

\subsection{Experimental setup}

\subsubsection{TF-IDF}
As mentioned earlier, we used two pre-processing routines to construct the TF-IDF matrices. The use of the complete pre-processing steps on the 15,966 sentences of the training set resulted in a matrix with 8,773 terms (columns) and is called ''TF-IDF Small''. In the case of simple pre-processing steps, the resulting matrix contained 15,004 terms, considering the same training set. This matrix is referred as ''TF-IDF Large'' and presents a sizable dimensionality increase in comparison to TF-IDF Small.

\subsubsection{Word Embeddings}
In order to measure the impact of embeddings trained on specialized (legal) documents, we provide a comparison of our method against pre-trained (off-the-shelf) embeddings. Those off-the shelf sets of embeddings were trained on Brazilian Portuguese documents \cite{Hartmann-2017}\footnote{http://nilc.icmc.usp.br/embeddings}. In this context, different embedding models, such as Word2Vec \cite{Mikolov-2013}, FastText \cite{Joulin-2016}, Glove \cite{Pennington-2014} and Wang2Vec (an extension of the Word2Vec published by Wang et al. \cite{Ling-2015}) have been tested with vector size equals to 100, 300 and 600 and the CBow and Skip-Gram techniques, except in the case of Glove, which has a different methodology that does not apply the alluded architectures. The above techniques were trained using a generic corpus, not specialized in the Law area. This method is called ``Generic Embedding''.

Additionally, we trained Word2Vec with a vocabulary similar to the classification dataset, using a corpus of 922,588 sentences extracted from standard operating prosecution procedures, also recorded in the PRO-MP system. For training the custom Word2Vec model with the PRO-MP dataset we used the package gensim\footnote{https://radimrehurek.com/gensim/}, applying CBoW and Skip-Gram architectures with window values of 5, 10, and 15, minimum word count equal to 2, 5 and 10 and vector size of 100, 300 and 600, in order to allow the comparison with the previous pre-trained models performance with the same vector size. This method is called ``Specialized Embedding''.

\subsubsection{Hyper Parameter Optimization}
We applied random search (using the training set) in order to optimize the parameters of the tested models, with a number of sampled parameter settings equals to 50. Due to computational limitations, the search was performed in a narrowed variety of parameter levels, applying 5-fold cross validation for the less complex models (linear classifiers and boosted trees) and 2-fold random permutation cross-validation for neural networks.

\subsubsection{Class Imbalance}
Finally, considering the unbalanced classes present in the dataset, we tested the impact of random balancing techniques. We employed a Random Under Sampler (RUS), which obtains a balanced random sample of classes, reducing all classes to the frequency of the smallest observed category. We did not use Random Over Sampler (ROS) because it would greatly increase the size of the dataset, making hyper-parameter optimization unfeasible (given the computational resources we had available for our research). The RUS technique applied to our dataset reduced the amount of data to 4,140 while the ROS technique would increase it to 97,110 observations.

\subsection{Results}

\subsubsection{Embeddings}
In order to determine the best parameter configurations of Word2Vec for the specialized embeddings, we adjusted several models and applied each resulting embeddings to a simple classifier to analyse the performance obtained. Thus, 54 models were created partitioning the training set into two sets in the proportion of 80/20. We then trained a SVM for each Word2Vec configuration in the larger partition, retrieving the accuracy and F1-Score on the smaller partition. The sentence representation was obtained as the mean vector of all word embeddings that composed the sentence. Considering that the best results were observed with vector size of 600, we present in Table \ref{table:specialized} the accuracy and F1-Score varying the other parameters and keeping the vector size equal to 600. The best results were achieved using Skip-Gram architecture, minimum word count equal to 5 and window value of 10.

\begin{table}[!ht]
  \caption{Results of SVM classifier using different configurations of the Specialized Embeddings. In bold the best F1-Score overall.}
  
  \label{table:specialized}
  \begin{tabular}{|c|c|c|c|c|} \hline
  
  & \textbf{Minimum} & \textbf{Window} & \multirow{2}{*}{\textbf{Accuracy}} & \multirow{2}{*}{\textbf{F1-Score}} \\
  & \textbf{Count} & \textbf{Size}  & & \\ \hline
  
  \parbox[t]{2mm}{
  \multirow{9}{*}{\rotatebox[origin=c]{90}{CBoW}}} &
  \multicolumn{1}{c|}{
  \multirow{3}{*}{2}} & 5  & 0.832 & 0.726 \\ \cline{3-5} &
                      & 10 & 0.841 & 0.749 \\ \cline{3-5} &
                      & 20 & 0.832 & 0.730 \\ \cline{2-5} &
  \multirow{3}{*}{5}  & 5  & 0.828 & 0.727 \\ \cline{3-5} &
                      & 10 & 0.842 & 0.747 \\ \cline{3-5} &
                      & 20 & 0.826 & 0.724 \\ \cline{2-5} &
  \multirow{3}{*}{10} & 5  & 0.825 & 0.715 \\ \cline{3-5} &
                      & 10 & 0.833 & 0.733 \\ \cline{3-5} &
                      & 20 & 0.831 & 0.736 \\ \hline \hline
  \parbox[t]{2mm}{
  \multirow{9}{*}{\rotatebox[origin=c]{90}{Skip-Gram}}} &
  \multicolumn{1}{c|}{
  \multirow{3}{*}{2}} & 5  & 0.868 & 0.793          \\ \cline{3-5} &
                      & 10 & 0.865 & 0.786          \\ \cline{3-5} &
                      & 20 & 0.864 & 0.782          \\ \cline{2-5} &
  \multirow{3}{*}{5}  & 5  & 0.865 & 0.783          \\ \cline{3-5} &
                      & 10 & 0.873 & \textbf{0.799} \\ \cline{3-5} &
                      & 20 & 0.870 & 0.793          \\ \cline{2-5} &
  \multirow{3}{*}{10} & 5  & 0.862 & 0.775          \\ \cline{3-5} &
                      & 10 & 0.867 & 0.786          \\ \cline{3-5} &
                      & 20 & 0.872 & 0.796          \\ \hline
  
  \end{tabular}
  
\end{table}

A similar test was performed on the generic embeddings. Again it was verified that the use of vector size equal to 600 showed improvement in the classifier performance. The results varying other parameters are presented in Table \ref{table:generic}, with best F1-Score achieved using the Wang2Vec technique, vector size equal to 600 and Skip-Gram. We note that the good performance of Wang2Vec in comparison to the other generic embeddings is in agreement with the results reported in Hartmann et al. (2017) \cite {Hartmann-2017}, in the sense that Wang2Vec presents good performance in a wide variety of NLP tasks.

\begin{table}[!ht]
  \caption{Results of SVM classifier using different configurations of the Generic Embeddings. In bold the best F1-Score overall.}
  
  \label{table:generic}
  \begin{tabular}{|l|c|c|c|} \hline
  
  \textbf{Model} & \textbf{Technique} & \textbf{Accuracy} & \textbf{F1-Score} \\ 
  \hline
  \multirow{2}{*}{Word2Vec} & CBow      & 0.796 & 0.684 \\ \cline{2-4}
  \multicolumn{1}{ |c| }{}
  & Skip-Gram & 0.807 & 0.705 \\ \cline{2-4}
  \hline
  \multirow{2}{*}{Wang2Vec} & CBow      & 0.820 & 0.721 \\ \cline{2-4}
  \multicolumn{1}{ |c| }{}
  & Skip-Gram & 0.844 & \textbf{0.762} \\ \cline{2-4}
  \hline
  \multirow{2}{*}{FastText} & CBow      & 0.803 & 0.708 \\ \cline{2-4}
  \multicolumn{1}{ |c| }{}
  & Skip-Gram & 0.830 & 0.744 \\ \cline{2-4}
  \hline
  \multirow{1}{*}{Glove} & -         & 0.827 & 0.738 \\ \cline{2-4}
  \hline
  
  \end{tabular}
  
\end{table}

\subsubsection{Linear Classifiers}
After setting the feature configurations, we trained all the aforementioned classifiers and measured the accuracy and F1-Score in the test set. The following table summarize the test results obtained by linear models trained using the original training dataset as well as the under-sampled (RUS) one.

\begin{table}[!ht]
  \caption{Results of Linear Classifiers in the Test Set. In bold the best F1-Score for each classifier.}
  \label{table:simpleclass}
  \begin{tabular}{|c|l|c|c|c|}
  \hline
  \multicolumn{1}{|c|}{} & \textbf{Configuration} & \textbf{Feature} & \textbf{Accuracy} & \textbf{F1-Score} \\ 
  \hline
  \parbox[t]{2mm}{
  \multirow{8}{*}{\rotatebox[origin=c]{90}{\textbf{Logistic Regression}}}} &
  \multicolumn{1}{c|}{
  \multirow{2}{*}{Small TF-IDF}} & RUS      & 0.80 & 0.76 \\ \cline{3-5} &
                                 & Original & 0.88 & \textbf{0.83} \\ \cline{2-5}  &
  \multirow{2}{*}{Large TF-IDF}  & RUS      & 0.77 & 0.71 \\ \cline{3-5} &
                                 & Original & 0.87 & 0.82 \\ \cline{2-5} &
  \multirow{1}{*}{Generic}       & RUS      & 0.79 & 0.73 \\ \cline{3-5} & 
                  Embeddings     & Original & 0.86 & 0.80 \\ \cline{2-5} &
  \multirow{1}{*}{Specialized}   & RUS      & 0.83 & 0.78 \\ \cline{3-5} & 
                  Embeddings     & Original & 0.87 & 0.80 \\ \hline \hline
  \parbox[t]{2mm}{
  \multirow{8}{*}{\rotatebox[origin=c]{90}{\textbf{SVM}}}} &
  \multicolumn{1}{c|}{
  \multirow{2}{*}{Small TF-IDF}} & RUS      & 0.79 & 0.75 \\ \cline{3-5} &
                                 & Original & 0.87 & \textbf{0.83} \\ \cline{2-5}  &
  \multirow{2}{*}{Large TF-IDF}  & RUS      & 0.77 & 0.72 \\ \cline{3-5} &
                                 & Original & 0.87 & 0.82 \\ \cline{2-5} &
  \multirow{1}{*}{Generic}       & RUS      & 0.77 & 0.71 \\ \cline{3-5} & 
                  Embeddings     & Original & 0.85 & 0.79 \\ \cline{2-5} &
  \multirow{1}{*}{Specialized}   & RUS      & 0.81 & 0.76 \\ \cline{3-5} & 
                  Embeddings     & Original & 0.88 & 0.82 \\ \hline
  \end{tabular}
\end{table}

From Table \ref{table:simpleclass} we can observe that the linear classifiers performed well, especially associated with the textual representation using TF-IDF matrices. In general, RUS did not bring any improvement when compared to the use of the original dataset. Sentence representation techniques using average word embeddings presented reasonable results, but were inferior to TF-IDF representations. In this regard, there was better performance using specialized embeddings compared to generic embeddings. Better performance was also observed using more qualified pre-processing in the creation of TF-IDF matrices.

\subsubsection{Boosted Trees}
The boosted tree classifiers presented in Table \ref{table:boostedtrees} under-performed the linear classifiers. Regarding the behavior in general, many similarities were observed, such as better performance using TF-IDF representations and worse performance using average embeddings to represente a sentence. There was also a drop in performance for the RUS setup and generic embeddings underperformed the specialized ones. The major difference in behavior was in relation to Gradient Boosting, which presented better results with TF-IDF matrix using simpler pre-processing steps.

\begin{table}[!ht]
  \caption{Results of Boosted Trees in the Test Set. In bold the best F1-Score for each classifier.}
  
  \label{table:boostedtrees}
  \begin{tabular}{|c|l|c|c|c|}
  \hline
  \multicolumn{1}{|c|}{} & \textbf{Configuration} & \textbf{Feature} & \textbf{Accuracy} & \textbf{F1-Score} \\ 
  \hline
  \parbox[t]{2mm}{
  \multirow{8}{*}{\rotatebox[origin=c]{90}{\textbf{Random Forest}}}} &
  \multicolumn{1}{c|}{
  \multirow{2}{*}{Small TF-IDF}} & RUS      & 0.79 & 0.73 \\ \cline{3-5} &
                                 & Original & 0.87 & \textbf{0.80} \\ \cline{2-5}  &
  \multirow{2}{*}{Large TF-IDF}  & RUS      & 0.76 & 0.68 \\ \cline{3-5} &
                                 & Original & 0.84 & 0.77 \\ \cline{2-5} &
  \multirow{1}{*}{Generic}       & RUS      & 0.63 & 0.56 \\ \cline{3-5} & 
                  Embeddings     & Original & 0.72 & 0.62 \\ \cline{2-5} &
  \multirow{1}{*}{Specialized}   & RUS      & 0.74 & 0.63 \\ \cline{3-5} & 
                  Embeddings     & Original & 0.79 & 0.68 \\ \hline \hline
  \parbox[t]{2mm}{
  \multirow{8}{*}{\rotatebox[origin=c]{90}{\textbf{Gradient Boosting}}}} &
  \multicolumn{1}{c|}{
  \multirow{2}{*}{Small TF-IDF}} & RUS      & 0.79 & 0.72 \\ \cline{3-5} &
                                 & Original & 0.84 & 0.76\\ \cline{2-5}  &
  \multirow{2}{*}{Large TF-IDF}  & RUS      & 0.77 & 0.70 \\ \cline{3-5} &
                                 & Original & 0.85 & \textbf{0.77} \\ \cline{2-5} &
  \multirow{1}{*}{Generic}       & RUS      & 0.71 & 0.65 \\ \cline{3-5} & 
                  Embeddings     & Original & 0.80 & 0.70 \\ \cline{2-5} &
  \multirow{1}{*}{Specialized}   & RUS      & 0.77 & 0.71 \\ \cline{3-5} & 
                  Embeddings     & Original & 0.84 & 0.75 \\ \hline
  \end{tabular}
\end{table}

\subsubsection{Neural Networks}
Finally, Table \ref{table:neuralnets} presents the results of the neural network architectures, all used with an embedding layer for sentence representation. 

It is worth noticing that five out of the six different combinations of neural net architecture/embeddings showed better F1-Score performance compared to their SVM equivalents. In all cases, the use of RUS decreased the classifier performance. The use of generic embeddings also led to poorer performance in relation to specialized embeddings. The first structure that combines a convolutional layer with max pooling was based on the work of Silva et al. (2018) \cite {Silva-2018}, with parameters reconfigured according to the results of the random search performed in our data. We highlight the superior performance of the RNN models (LSTM and GRU), the first being the best classifier obtained in this work, with an F1-Score of 0.8549 and accuracy of 0.9047.

\begin{table}[!ht]
  \caption{Results of Neural Networks in the Test Set. In bold the best F1-Score for each classifier.}
  \label{table:neuralnets}
  \begin{tabular}{|c|l|c|c|c|}
  \hline
  \multicolumn{1}{|c|}{} & \textbf{Configuration} & \textbf{Feature} & \textbf{Accuracy} & \textbf{F1-Score} \\ 
  \hline
  \parbox[t]{2mm}{
  \multirow{4}{*}{\rotatebox[origin=c]{90}{\textbf{CNN+Max}}}} &
  \multicolumn{1}{c|}{
  \multirow{1}{*}{Generic}}    & RUS      & 0.75 & 0.71 \\ \cline{3-5} & 
                  Embeddings   & Original & 0.87 & 0.80 \\ \cline{2-5} &
  \multirow{1}{*}{Specialized} & RUS      & 0.80 & 0.74 \\ \cline{3-5} & 
                 Embeddings    & Original & 0.88 & \textbf{0.82} \\ \hline \hline
  \parbox[t]{2mm}{
  \multirow{4}{*}{\rotatebox[origin=c]{90}{\textbf{LSTM}}}} &
  \multicolumn{1}{c|}{
  \multirow{1}{*}{Generic}}    & RUS      & 0.79 & 0.73 \\ \cline{3-5} & 
                  Embeddings   & Original & 0.88 & 0.83 \\ \cline{2-5} &
  \multirow{1}{*}{Specialized} & RUS      & 0.82 & 0.76 \\ \cline{3-5} & 
                  Embeddings   & Original & 0.90 & \underline{\textbf{0.85}} \\ \hline \hline
  \parbox[t]{2mm}{
  \multirow{4}{*}{\rotatebox[origin=c]{90}{\textbf{GRU}}}} &
  \multicolumn{1}{c|}{
  \multirow{1}{*}{Generic}}    & RUS      & 0.82 & 0.76 \\ \cline{3-5} & 
                  Embeddings   & Original & 0.89 & 0.83 \\ \cline{2-5} &
  \multirow{1}{*}{Specialized} & RUS      & 0.84 & 0.78 \\ \cline{3-5} & 
                  Embeddings   & Original & 0.89 & \textbf{0.84} \\ \hline \hline
  \end{tabular}
\end{table}

\begin{figure}[h!]%
  \centering
  \hspace*{-.1in}
  \includegraphics[width=9cm]{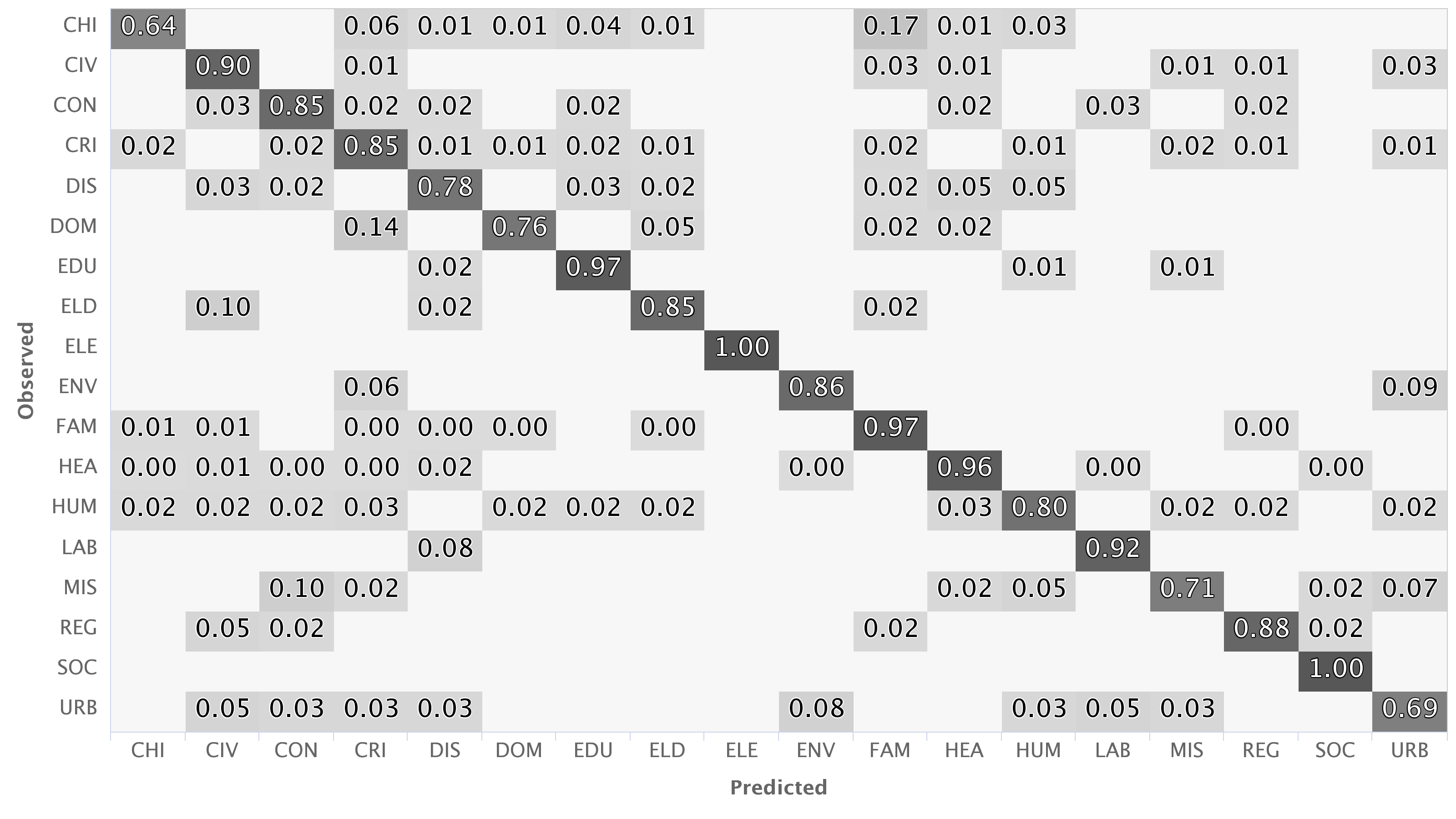}
  \caption{Confusion matrix produced by the LSTM using Specialized Embeddings}
  \label{fig:cm_prop}
\end{figure}

Considering that the best results were obtained with an LSTM associated with specialized embedding representation, we present the confusion matrix of this classifier in the test set, representing the eighteen fields of law analyzed in this study (Figure \ref{fig:cm_prop}). 

With the detailing of the classes it is possible to verify that the highest indices of confusion of the model are associated with the classes "Urban Planning Law" (URB) and "Child and Youth Law" (CHI), while the most precise values are concentrated in the "Electoral Law" (ELE) and "Social Security Law" (SOC) areas, both with no wrong predicted cases in the test set. Another aspect identified was the high degree of confusion between the "Child and Youth Law" (CHI) and "Family Law" (FAM) classes, which can be partly explained by the fact that the vocabularies of these areas are often correlated, being complex even for a human being to differentiate and correctly categorize such type of petition. This problem also occurs with "Criminal Law" (CRI) and "Domestic Violence Law" (DOM) fields. In general, we can observe that the use of recurrent neural networks presents promising results and can be applied in other existing problems in the Public Prosecutor's Office of Paraná.

\section{Conclusions and Future Work}
\label{sec:conclusion}

In the present study it was verified that the construction of text classifiers applied to the Law area is possible, although laborious. Data pre-processing is an essential preliminary step to some techniques, being associated with the quality of the information and, consequently, the accuracy and reliability of the classifier.

Especially in the case studied, it was possible to verify that different forms of word representations can lead to acceptable results and it is important to choose the best method considering the cost (human and computational) associated with pre-processing. Also, matching the representation to the right family of models is very important. It was possible to identify that despite the use of ``off the shelf'' POS tagging and lemmatization, it still led to good results. It was also noted that a more superficial data cleaning with the application of semantic techniques achieve similar or better results as an extensive pre-processing, although they may be more laborious in the training and parametric selection steps. Still on the subject, it was possible to identify that the use of specialized vocabularies present considerable gains in relation to the use of generic ones (despite the smaller vocabulary size).

Regarding class imbalance, we found out that the application of random balancing techniques led to poorer results (in terms of F1-Score) for all the classifiers evaluated in this paper. We emphasize that in that scenario, even the application of random search did not lead to any significant performance improvement. The poorer performance was possibly due to the decrease in the size of the training corpus caused by under-sampling. For future work, consideration should be given to using over-sampling to identify whether re-balancing without decreasing the training set size has the potential to improve performance.

As for the different classification models evaluated in this paper, it was observed that both, the combination of simple representations with simple classifiers, as well as the combination of word embeddings with more complex models (neural networks) led to good results. A lower performance was observed with the boosted tree classifiers in relation to the others models evaluated in this paper. We will consider the possibility of testing other similar structures in future work. Recurrent neural networks obtained the best results in the experiments (especially LSTM). Other architectures showed no gain in the evaluated metrics, but still were superior to the simpler models.

It is worth noticing that the best accuracy observed in our experiments (above 90\%) largely surpasses the human accuracy for the same task (72\%) observed in preliminary studies. Thus, we can see an immediate benefit in terms of information quality with the implementation of the model as an automation tool in the petition management system (PRO-MP). In terms of usability, on inference time, we suggest presenting to the user the 5 categories with highest confidence (for a given document) and let the user manually disambiguate as the accuracy $@5$ was 99\%.

Regarding the limitations, it is important to highlight that, despite the validation steps performed in the beginning of this study, some incorrectly labeled observations are still present in the dataset. In addition, some fields of law currently have overlapping competencies, so that, depending on the nature of the description, it is not possible, even for a human being, to assign a single unambiguous label to every petition. Despite these limitations, we find the resulting model appropriate and useful for the MPPR routine with preliminary tests showing an almost immediate prediction to the user when applying simple textual pre-processing combined with the LSTM neural network.

Future works are intended to deepen the present study, in order to enable the integration with a model capable of identifying the most relevant words to the predicted class, inspired by the work of Arras et al. (2017) \cite{Arras-2016}. Thereby, it will be possible to identify, from the description of a case, not only the field of law but associated keywords as well. We also intend to update some of the methodologies used in order to obtain a contextualized sentence representation, applying techniques that make use of language models and transformers architectures, such as the recently developed BERT technique. Finally, we highlight that the insertion of the current model in the electronic system of records of the Public Prosecutor's Office will guarantee the improvement of the work flow and statistics generated by the institution, reverting in concrete benefits to the population as a whole.

\section*{Acknowledgements}
We gratefully acknowledge the contributions of the following MPPR colleagues that helped us in this research: Monica Louise de Azevedo, Admilson Aparecido Garcia Buzinaro, Claudio Fernando Weigratz Tavares, Elaine Mara Vistuba Kawa, Denise Ratmann Arruda Colin, Murilo Cezar Soares e Silva and Marcos Bittencourt Fowler.

\bibliographystyle{unsrt}
\bibliography{rbef}

\end{document}